# Making it normal for 'new' enrollments: Effect of institutional and pandemic influence on selecting an engineering institution under the COVID-19 pandemic situation

Corresponding author (*)
**Prashant Mahajan (*)**
R. C. Patel Institute of Technology, Shirpur.
registrar@rcpit.ac.in
ORCiD ID: 0000-0002-5761-5757

**Vaishali Patil**
RCPET's, Institute of Management Research and Development, Shirpur
imrd.director@gmail.com

**Abstract**

The COVID-19 pandemic has forced Indian engineering institutions (EIs) to bring their previous half-shut shades completely down. Fetching new admissions to EI campuses during the pandemic has become a 'now or never' situation for EIs. During crisis situations, institutions have struggled to return to the normal track. The pandemic has drastically changed students' behavior and family preferences due to mental stress and the emotional life attached to it. Consequently, it becomes a prerequisite, and emergencies need to examine the choice characteristics influencing the selection of EI during the COVID-19 pandemic situation.

The purpose of this study is to critically examine institutional influence and pandemic influence due to COVID-19 that affects students' choice about an engineering institution (EI) and consequently to explore relationships between institutional and pandemic influence. The findings of this quantitative research, conducted through a self-reported survey, have revealed that institutional and pandemic influence have governed EI choice under the COVID-19 pandemic. Second, pandemic influence is positively affected by institutional influence. The study demonstrated that EIs will have to reposition themselves to normalize pandemic influence by tuning institutional characteristics that regulate situational influence and new enrollments. It can be yardstick for policy makers to attract new enrollments under pandemic situations.

**Keywords**

Engineering education, choice characteristics, institutional influence, pandemic influence, suitability under the COVID-19 and COVID-19 pandemic situations.

## Introduction

Worldwide, engineering education is viewed as a career of progressive growth that has the potential to shape human skills (Blom & Saeki, 2011), social and quality of life (Rojewski, 2002), the economy of the country (Cebr, 2016) and the overall development of the country (Downey & Lucena, 2005). Thus, engineering education has proven to be a key factor for the sustainable and profitable development of society. It encourages global competitiveness through engineering inventions for the benefit of society at large. However, although the demand for engineers remains relatively high throughout the world, there are few aspirants willing to pursue engineering education. Creating an upswing for interest and fondness that makes students inclusive of engineering studies itself has been a challenge. Reports on engineering education about declining enrollments (AICTE New Delhi, 2021) in the context of India and diminishing interest and trends worldwide (UNESCO, 2010) have signaled a warning bell for the overall development of hi-tech society. In India, the gap between available seats (capacity) at the entry level and actual admissions in degree engineering is widening year by year, leaving approximately 5.9 lacs seats vacant in 2019-2020. All India Council for Technical Education, New Delhi, an apex body for governing technical education, indicated that approximately 45% of seats remained vacant in the 2019-2020 academic year, which was noticed to be 38% in 2012-2013. Most of the studies have verified that the situation is due to problems pertaining to awareness, attraction, recognition of needs and service offers (Kamokoty et al., 2015; Upadhayay and Vrat, 2017).

Selecting an institute, as acknowledged by previous literature, is a subtle and complex phenomenon (Hossler et al., 1989a) that involves a multifaceted and inconsistent set of institutional influencing characteristics (Obermeit, 2012) (Chapman, 1981). It implicates a challenging progression for institutions as well as aspirant students (Hemsley-Brown & Oplatka, 2015b) and requires greater efficiency and effectiveness to make a concluding decision. Decisions regarding 'institutional choice' can change students' lives forever (Iloh, 2019) and the performance of institutions. Selecting an engineering institution (EI) has not received much consideration but is practically missing in the literature, as the research drift appeared to be inclined towards general higher education addressing psychology, sociology, and economics disciplines (Paulsen, 1990). Today, most EIs in India with lower enrollments are in vilest positions due to the absence of practicing students' assessment in regard to their needs. Engineering education is highly contrasted with respect to the multidimensional thoughts of students and institutional influence related to the quality of staff and teaching-learning, infrastructure and facilities, course value and delivery, and outcome benefits.

*Statement of the problem*

There is certain evidence that higher education (HE) needs to be drastically reformed due to unforeseen situations or crises due to political and economic changes arising due to natural disasters (Schuh & Santos Laanan, 2006) and pandemics (Kim & Niederdeppe, 2013). In such a situation, HE institutions have struggled to return on the normal track. (Aristovnik et al., 2020) revealed that the COVID-19 pandemic has changed students' emotional and personal lives and has also changed their preferences and habits in regard to the selection of HE. The survey conducted by The International Association of Universities discovered that COVID-19 will affect future enrollment for upcoming academic years (IAU, 2020). Consequently, it becomes a prerequisite, and emergencies need to examine the choice characteristics influencing the selection of EI during the COVID-19 pandemic situation. Second, in such a pandemic situation, an examination of students' perceptions of choice characteristics holds great practical importance for policy makers and administrators of EIs.

*Objective of study*

As informed by the evidence and problems discussed above, the main objective of this study is to critically examine choice characteristics related to institutions and pandemics that influence students' choice for EI during the pandemic and consequently to explore relationships between institutional and pandemic influence arising due to COVID-19. The above objective is underpinned by the following research questions referring to the selection of an EI during the COVID-19 pandemic.

1. What are the important institutional and pandemic influencing characteristics that have influenced potential students' decisions about selecting EI during the pandemic?
2. How are institutional influence and pandemic influence coupled?

## Literature review

This study embraces a systematic review (Bearman et al., 2012) that progressed gradually through extensive searching, selecting and integrating literature that has explored the evolution and influence of choice characteristics responsible for the selection of an institution. The literature review revealed that the institute choice process has reformed over time in accordance with ecological changes (Jackson, 1988), informed awareness and understanding of institutional facilities (Nora & Cabrera, 1992). To make a pathway for prospective students, HE institutions should understand who students are and what they expect from them and how their expectations can be met by educational offers (Han, 2014). Hemsley-Brown & Oplatka (2015) learned that despite ample literature, there is no assured list of choice characteristics that influence and confirm students picking up a specific institute. The following section describes at length the characteristics linked to institutional influence that are accountable for students' choice decisions.

*Institutional influence*

Institutional influence is a set of characteristics that magnetizes prospective students towards institutions. These characteristics are clustered on financial vs nonfinancial offers, academic vs nonacademic facilities and services and tangible vs intangible factors (Hossler et al., 1989b) (Yamamoto, 2006), which are reviewed below.

*Proximity to hometown*

Proximity relates to the nearness of the hometown from the institution. Being close to an institution is a significant factor for students in selecting an institution (Turley, 2009). It also increases the chance of acceptance for the particular institution (López Turley, 2009), as distance travel is associated with cost, time and efforts (Chapman, 1981). In the case of engineering study, due to a heavy workload, it can provide extended hours for study at home and enough time for social and other activities if EIs are situated near students' hometowns.

*Location and locality*

Location and locality are structures of ambient conditions, speciousness and functional accessibility (Bitner, 1992) and are swaying characteristics in making institutional choices (Gibbs & Knapp, 2012). Location gives impression of institute's site and its connectivity from hometown, while locality refers to culture, amenities, and facilities available in surrounding place wherein the institution is located. Overall, it is credited with suitability, vicinity, attractiveness, accessibility, cost-effectiveness, safety and security (Hannagan, 1992; Kotler & Fox, 1995)

*Image and reputation*

Image and reputation in public minds plays a significant role in differentiating institutions (Imenda et al., 2004) and is measured as one of the topmost characteristics influencing institution choice (Briggs, 2006) (Wadhwa, 2016). It is composed of a spectrum of small reputes, such as academic and nonacademic characteristics belonging to institutions (Lafuente-Ruiz-de-Sabando et al., 2018). In the review of the literature (Hemsley-Brown & Oplatka, 2015c) and in most of the research (Maringe, 2006b), the image and reputation of institutions provide the first impression that embosses decision makers in minds, even if nobody is confronted with institutions.

*Faculty profile*

Faculty profiles in terms of their qualifications, skills, competency and experience (Imenda et al., 2004) exert a significant influence on students (Soutar & Turner, 2002) (Mazzarol & Soutar, 2002). Faculty ought to be profiled with high-quality teaching (Woolnough, 1994) and should be a well designer (De Courcy, 1987). Similarly, they should be well-inspired, well informed, passionate, open minded, and responsive (Voss et al., 2007) to transform knowledge and to assist students in real-world exposure (Bhattacharya, 2004). (Magnell et al., 2017) mentioned the importance of faculty attitudes in assisting students in availing engineering curricula.

*Alumni image*

Alumni are the tangible outcome of institutions, and hence, alumni concerns are important criteria in measuring the performance of EIs. Alumni achievements are often exploited to exemplify the importance, eminence and image of institutions (Saunders-Smits & de Graaff, 2012) and criteria for selecting an institution (Ho & Hung, 2008). Historically, alumni images with economic, market and social standing at all times have added glory to the reputation of their institutions and hence have become benchmarking standards for prospective students (Pucciarelli & Kaplan, 2016).

*Campus placements*

Employment prospects are the potential outcomes and benefits that prospective students and their families seek against time, effort and money invested in HE institutions (Hemsley-Brown & Oplatka, 2015c) (Maringe, 2006a). The transition from education to employment is the straightforward motive of every student opting engineering study (Baytiyeh & Naja, 2012) and has been verified to be one of the most influential characteristics in making institutional choices (Malgwi et al., 2005). Most premium EIs have a series of campus placement activities dealing with students' employment and upholding alliances between industry and academia. It has a major role in boosting employability skills (Markes, 2006) and accelerating industry-academia connections (Baytiyeh & Naja, 2012) to create employment opportunities for engineering students.

*Quality Education*

Quality of education is a prime, discriminating, and prominent influencing characteristic of EI consigned to stay ahead in a competitive market and to make a place in the minds of stakeholders. Several studies (Pandi et al., 2014) (Sakthivel & Raju, 2006a) (Sayeda et al., 2010) have emphasized the importance of quality education in regard to the holistic development of institutions and in making institute choice decisions for students (Kallio, 1995) (Mourad, 2011). Several items, such as academic standards, industry linkages, and campus placements, contribute to the quality of education (Mahajan et al., 2014). Furthermore, for some researchers, it implies course delivery (Trum, 1992), infrastructure facilities (Sayeda et al., 2010), faculty (Gambhir et al., 2013), quality services (Viswanadhan, 2009), and academic and nonacademic concerns (Jain et al., 2013) (Owlia & Aspinwall, 1998). Overall, it has a two-fold effect in terms of tangible and intangible outcomes (Natarajan, 2009).

*Infrastructure and facilities*

The importance of infrastructure and facilities is mentioned in numerous studies, such as (Nyaribo et al., 2012) (Sahu et al., 2013) (Price et al., 2003). It consists of buildings, equipment, infrastructure and amenities that are tangible possessions reflecting the capacity of institutions that streamline the performance of curriculum delivery (Palmer, 2003). It can provide love-at-first-sight and becomes on-the-spot evidence for prospective students (Kotler et al., 2002). Delivering a curriculum without the existence of infrastructural assets and facilities is not possible for EIs, as delivery is more technical in nature.

*Safety and security*

Safety on the institute campus means the provisions made about residential, physical health, and life concerns to ensure the wellbeing of students (Ai et al., 2018), whereas security, as a broad term, covers human rights, emotions and cultural values associated with students (Calitz et al., 2020)**.** Studies such as (Elliott & Healy, 2001) (Peters, 2018) have exposed that students contemplate it based on wellbeing and humanize culture, whereas (Calitz et al., 2020) revealed that it traditions allied with decisions about the selection of institutions. The students feel comfortable with the health services, emergency and situational provisions delivered by the institute (Sakthivel & Raju, 2006b).

*Curriculum delivery*

In engineering education, curriculum delivery is the most influential characteristic and is found to be the first priority in selecting an EI in most studies, such as (Moogan & Baron, 2003). It is associated with execution of a planned pedagogy supported by intangible services and tangible facilities that ensures continuous transfer of knowledge (Case et al., 2016). Engineering institutions can bring glory to the institute if delivered as per the needs but can take vilest situations if not handled properly. Curriculum delivery involves mix-up modes such as online (Alawamleh et al., 2020), hybrid (Tan, 2020) (Sia & Adamu, 2020) and regular onsite delivery depending on the situational crises. Although all have their own advantages and disadvantages in regard to the involvement of theory vs practical, technology vs human, and competency skills achieved, the degree to which it facilitates accessing, practicing and implementing knowledge is more important (Shay, 2014). To attract enrollments, delivery of engineering curriculum is to be considered a backbone that transforms engineering knowledge into practical applications (Hemmo & Love, 2008).

*Value for money*

Value for money is an intangible characteristic and deemed to be an anxiety for students that influences their selection of institutions. In engineering studies, the nature of costs is differential and includes tuition, travel, residential and food costs, and day-to-day academic costs, which are more expensive than other higher education methods. Some studies have exhibited the cost of education as a package of rewarding value benefit entailing, value and quality (Ivy, 2008) (Joseph et al., 2005), time and effort (Kotler & Fox, 1985), effort and opportunity (Wu et al., 2020). The degree of engineering, employment opportunities, skills gained, and social status are the foreseen values for students against their financial investment.

*Pandemic influence*

Pandemic influence is. Pandemic influence referring to this study is all about pandemic situation triggered due to Covid-19 pandemic situation occurred due to corona virus. It is an external influence that affects customers' behavior due to psychological perceptions about the situation (Belk, 1975). COVID-19 is a disease triggered by coronaviruses first discovered in December 2019 that causes respiratory illness spread though small saliva in the form of droplets and aerosols arising out due to close human contacts (Ciotti et al., 2020). As indicated by the World Health Organization, physical and social distancing is the only credible way to constrain its spread. It has taken out higher education by storm and hence turns out to be the most challenging condition in the history of engineering education. A US-based study (Aucejo et al., 2020) showed that the influence of the COVID-19 pandemic on HE is extremely heterogeneous. In the past, during situational crises, (Rosenthal et al., 2014) emphasized appropriate curriculum delivery, and (Kim & Niederdeppe, 2013) suggested students' support system as an important factor in normalizing the situation and continuing pedagogy.

After unlocking pandemic restrictions in August 2020 in India, the admission process for engineering studies for new enrollments for the 2020-2021 academic year in the state of Maharashtra was completed in January 2021. EIs were able to commence academic sessions for newly joined students from February 2021, as per the guidelines (UGC, 2021a) and State Government norms, that restricted onsite pedagogy with a 50% batch size on a rotation basis. Meanwhile, there were many pros and cons about curriculum delivery under the COVID-19 pandemic situation. To some authors, online delivery is most suitable during the pandemic to continue education further (Gautam & Gautam, 2020; Liguori & Winkler, 2020). However, it has been adversely condemned for various reasons, such as technology availability, academic loss and ongoing interest (Bird et al., 2020) (Zia, 2020) (Tesar, 2020). Some authors have suggested hybrid/blended delivery (Rashid & Yadav, 2020; Sia & Adamu, 2020) as a solution to continuing pedagogy during the pandemic. (Aristovnik et al., 2020) revealed that the pandemic with emotional life has also affected certain behavioral characteristics in terms of their likings and preferences. The pandemic situation has stressed potential students to think more about better accessibility and fitness. Therefore, there is an urgent need for policy reforms that sustain the mental health and social emotions of students (WHO, 2020). (UNESCO, 2020) judges that education has to be redefined or reduced and replaced or enhanced to engage students, particularly to avoid academic, social and emotional loss. (Chadha et al., 2020), in a recent study on UK engineering students, articulated that there is more need to implement new reforms to ensure that engineering students should not go down its pathway.

Thus, the pandemic influence referring to this study is EIs' efforts and provisions for making engineering education sustainable by providing suitable facilities and support services that mitigate the impact of the pandemic on students' pedagogy by following government guidelines about social distancing.

*Research gap and significance of study*

Many researchers have notarized a variety of characteristics influencing selection decisions about institutions, originating due to different cultures and economic and social reforms, but all were administered under nonpandemic situations. Many researchers felt that students' behavior changed during the COVID-19 pandemic, and there is urgency to reposition the framework of policies, which demanded future research that urges exploring institutional choice characteristics and pandemic influence during the pandemic.

Moreover, there is no such research to date that provides knowledgeable relationships between students' perceptions of EI selection during the COVID-19 pandemic. The importance and timeliness of this study is boundless, as it aimed to explore radical changes that materialized in students' choice characteristics during the COVID-19 situation.

*Conceptual framework and hypothetical model*

The literature review has shown that choice decisions are based on attractive and beneficial offers made by institutions in regard to tangible facilities and intangible services. However, during the COVID-19 pandemic, the process of evaluating alternatives involves a more intellectual and meticulous screening of institutional characteristics and the external influence of the pandemic that determines the suitability and accessibility of educational services by following social distancing norms that restrain the infection and spread of the disease, COVID-19.

Based on the theoretical and conceptual framework as stated above and the specified objective of the study, the following hypothetical model (refer to Figure I) will stand for answering the research questions. The following null hypothesis is validated based on students' perceptions.

*Hypothesis ($H_0$)*

There is no significant relationship between students' perceptions of institutional influence and pandemic influence when selecting an engineering institution under the COVID-19 pandemic.

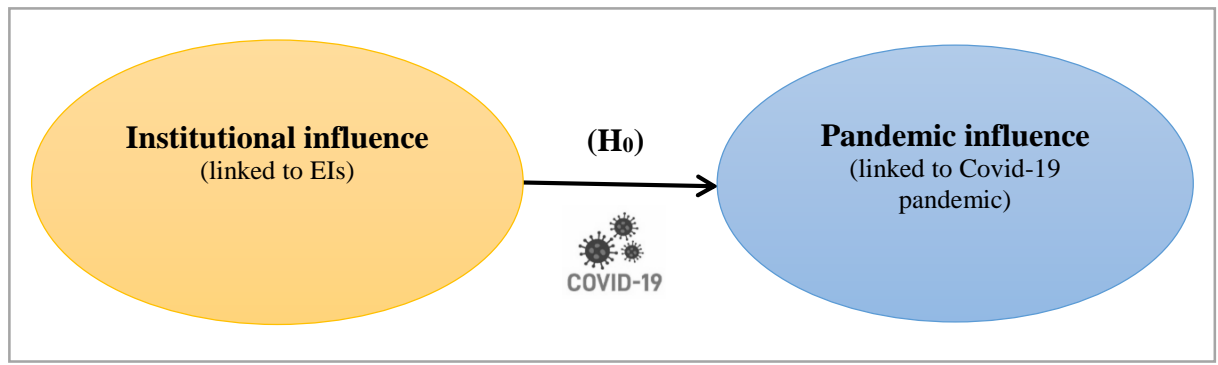

Figure I: Hypothetical path model of choice influencing characteristics under the COVID-19 pandemic
Source: Own

## Research Methodology

*Research design*

This study is marketing research about an educational dilemma associated with EI choice, particularly during the COVID-19 pandemic. A literature review aligned with the objective of this study has enabled this study to implement quantitative methods due to their ability to frame hypotheses (Borrego et al., 2009), capabilities to operate on multivariate statistical data (Creswell & Creswell, 2017), ability to analyze relationships with definiteness (Creswell, 2012b) and reliability (Steckler et al., 1992) and success in educational research (Tight, 2015).

The judgment of what students truly receive from the HE service against their expectations is often based on the evaluation of students' perceived experience (Yelkur, 2000). Therefore, this study has considered students' perceived experience as being primary customers of HE (Maringe & Gibbs, 1989). Primary data are collected using a survey method that is most suitable as per (Kotler et al., 2016) for collecting preferences and choices from a large number of responses. Students who recently enrolled in EIs during the COVID-19 pandemic situation were selected as a population of this study. Purposive sampling has been chosen decisively because of the knowledge and judgment of researchers (Creswell & Creswell, 2017), the special situation (Neuman, 2013) and the investigation of new issues (Etikan et al., 2016) about 'EI choice' during the COVID-19 pandemic.

The study is set to report the perceived experience of students about their pathway to an engineering institution during the pandemic. EIs offering degree courses in engineering and technology situated in the North Maharashtra region of India were chosen as the sampling frame of this study. The admission process for the first-year degree engineering program for 2020-2021 was conducted under the control of the competent authority of Maharashtra State, India, and ended on January 31, 2021. Sampling units consisting of newly joined students from a batch of academic years 2020-201 who recently experienced their EI selection process under the COVID-19 pandemic were chosen from the sampling frames. A total of 4300 e-mail addresses of students representing academic batches of 2020-2021 from 39 units of sampling frames (EIs) were collected on the researcher's request through e-mail during February 2021. To make students more responsive, a self-report survey (Kolb, 2008) was conducted over the internet via the Google Form tool during February 1-15, 2021.

During the pandemic, a self-reported survey was very useful, as it avoided direct contacts with the respondents during the pandemic but at the same time ensured its reach to the expected respondents (students). This method also assisted in receiving responses quickly by providing respondents with better flexibility in time and place and avoided researcher bias. The survey received 922 responses overall at a response rate of 21% in mid-February 2021. (Creswell, 2012a) recommended a sample size of at least 20 samples per variable. This study included twelve independent variables with 922 valid responses. A sample size of 922 for assessing twelve variables, which derives 77 samples per variable, is sufficiently defensible against the traditional arbitrary ratio of 20:1 (Maxwell Scott, 2000).

*Scale design and data collection*

A quantitative survey is administered with a list of structured closed-ended questionnaires that were developed on the guidelines provided by (Cohen et al., 2007) and (Ary et al., 2010). The questionnaire was initiated with an introductory part, Section I, explaining the purpose and importance of the study. Section II presents three questions on students' personal characteristics, such as gender, social class, and native place. Section III was associated with influencing institutional characteristics, which are evidenced under a literature review and recommended by academic experts. The section encompassed twelve choice characteristics that are continuous in nature and influenced students' decisions about the selection of their EIs. In this regard, students are asked to rate on a Likert scale (1 to 5) their perceived experience with proximity, location and locality, image and reputation, faculty profile, alumni records, campus placements, quality education, infrastructure and facilities, safety and security, curriculum delivery, value for money and suitability under the COVID-19 pandemic situation. Before entering the actual survey, the validity and reliability of the questionnaire was tested through pilot testing (Kenneth, 2005) with few samples selected from sampling units to understand its language and sequence of questions and was noticed to be suitable for conducting the actual survey.

## Data analysis and statistical results

Making institutional choices under the COVID-19 pandemic situation is a new difficult encounter for students. In such a situation where influencing characters ruling choice decisions and their connections are unfamiliar, data examination is executed by a two-step approach (Anderson & Gerbing, 1988). To determine the relationship between influencing characteristics associated with EIs and pandemic influence, exploratory factor analysis (EFA) and structural equation modeling (SEM) were performed. EFA is first performed to develop constructs (latent variables) from item scales (observed variables), followed by confirmatory factor analysis (CFA) performing structural equation modelling (SEM) to predict the relationships between the extracted constructs (Byrne, 2013a). In the first stage, factor analysis by EFA was performed on twelve influencing characteristics that recognize importance in selecting an EI during the COVID-19 pandemic. The second stage incorporated CFA and SEM by developing a measurement and structural model that represents the best relationship between components extracted from EFA. The data were analyzed and analyzed with the techniques available in the statistical software SPSS 25.0 and AMOS 25.0. Before arriving at the EFA and SEM results, the statistical fitness of the data in terms of sample adequacy, reliability and validity are justified, as discussed below.

*Statistical fitness of data*

Reliability based on internal consistency was successfully validated by Cronbach's alpha, item-total correlation, and the split-half technique available in SPSS under reliability analysis (refer Table I). Values of Cronbach's alpha are above 0.6 for all scale items that have confirmed scales' internal consistency (Churchill Jr, 1979) and are best fit for the purpose (Nunnally & Bernstein, 1967). Next, corrected item-total correlations, which are noticed above 0.33, indicated good internal consistency of scales (Briggs & Cheek, 1986) and are found below 0.85, which proves no potential issues on multicollinearity (Kline, 2005). The split-half method successfully correlated half of the scale items with the other reaming half. For both parts, the value of the Spearman-Brown coefficient displayed the same value (0.93) within the parts, which expressed that the observed variables have more internal consistency with their latent variables (Ho, 2006). Composite reliability (CR) and average variance extracted (AVE) for each extracted latent variable derived from EFA were calculated (Refer Table I). The obtained values are well above the acceptable level of 0.7 (Fornell & Larcker, 1981) for CR and above 0.5 for AVE (Joseph et al., 1998). Last, Tukey's test was effective in detecting no additivity, which confirmed a sufficient estimate of power.

The scale items under this study signifying influencing choice characteristics about EIs are collected from rigorous analysis of the literature. In addition, academic EI experts have confirmed these influencing characteristics responsible for the inclusion of students in EIs. Factor loadings for all observed variables are well above 0.4, indicating that all twelve scale items are loaded strongly and significantly, confirming strong construct validity for their respective latent variables (refer Table I). Finally, no single scale items were noticed to have factor loadings above 0.4 across another construct (excluding own construct), which suggested that all scale items clarify sound discriminant validity (Ho, 2014) (Joseph et al., 1998). Because each scale item has loaded on only one latent variable, there is evidence of convergent and discriminant validity.

*Step I - Scale reduction and component extraction by EFA*

EFA proceeds to determine how and to what extent the observed variables are connected to their underlying component (latent variable) (Byrne, 2005). To start with EFA, all twelve choice characteristics have been processed with varimax rotation keeping the eigenvalue above 1.0 (Ho, 2006). Overall, twelve scale items have demonstrated a high level of potential for being factorized, with a Kaiser-Meyer-Olkin (KMO) measure of sampling adequacy at value 0.958 which is greater than required value (>0.5) as suggested by (Joseph et al., 2006). Furthermore, with the value of $\chi^2$= 7328.117 with $df$=66 significant at $p$<0.000, Bartlett's test of sphericity has shown creditable adequacy for factor analysis (Cerny & Kaiser, Henry, 1977).

EFA extracted two main components having common features within components; however, they were dissimilar across the components (refer to Table I). The first component was extracted from ten scale items (C1 to C10) accounting for 59.2 percent of the variance and is labeled 'institutional influence' (II), as all ten scale items represent traditional institutional characteristics that were usually accessed by students during nonpandemic situations for EI selection. Cronbach alpha for this component is 0.944. The second component explained 8.77 percent of the variance and exhibited an eigenvalue of 1.052 (above 1.0). It comprises two scale items (C11 and C12) symbolizing choice characteristics that influence potential students under the COVID-19 pandemic in directing EI choice decisions. This component is classified as 'pandemic influence' (PI). Cronbach's alpha for this component was 0.627, which was lower than that of the previous component due to the few item scales associated with it; however, the Cronbach's alpha was within acceptable limits (Tavakol & Dennick, 2011). Labeling of components is created on the type of scale items it houses and its relevance to the reviewed literature on institutional choice. The factor loading for the first extracted component ranged from 0.682 to 0.847, and for the second component, it ranged from 0.619 to 0.923, showing strong construct validity. The hypothetical path model was estimated to assess the explanatory power of all independent observed variables associated with the latent variables. Then, Step II proceeds to justify the strength and significance of the relationships by performing CFA and SEM, as discussed below (Table II, II and Figure II).

**Table I: EFA results with reliability and validity**

| Observed variables | | Mean | Corrected Item-Total correlation | Latent variables | |
|---|---|---|---|---|---|
| | | | | Component 1 | Component 2 |
| Choice characteristics | Code | | | | |
| Location and locality | C1 | 3.992 | 0.745 | 0.761 | -- |
| Image and reputation | C2 | 4.044 | 0.759 | 0.799 | -- |
| Faculty profile | C3 | 3.964 | 0.805 | 0.847 | -- |
| Alumni profile | C4 | 3.906 | 0.789 | 0.822 | -- |
| Campus placements | C5 | 3.979 | 0.769 | 0.825 | -- |
| Quality education | C6 | 3.937 | 0.785 | 0.809 | -- |
| Infrastructure and facility | C7 | 3.911 | 0.794 | 0.805 | -- |
| Safety and security | C8 | 3.972 | 0.783 | 0.798 | -- |
| Curriculum delivery | C9 | 3.929 | 0.789 | 0.794 | -- |
| Value for money | C10 | 3.764 | 0.677 | 0.682 | -- |
| Proximity | CC1 | 3.329 | 0.464 | -- | 0.923 |
| Suitability under Covid-19 | CC2 | 3.502 | 0.464 | -- | 0.619 |
| No. of scale items | | | | 10 | 2 |
| Eigen value | | | | 7.105 | 1.052 |

| | | | |
|---|---|---|---|
| % variance | | 54.369 | 13.603 |
| α based on standardized items | | 0.944 | 0.627 |
| Composite reliability | | 0.945 | 0.757 |
| AVE | | 0.633 | 0.618 |
| Component labeling | | Institutional influence (II) | Pandemic influence (PI) |

Notes: Extraction method: principal component analysis (PCA).
Rotation Method: Varimax with Kaiser Normalization.
Rotation converged in three iterations with extraction of two components.

*Step II – Executing the measurement model through CFA and SEM*

CFA and SEM were performed according to the guidelines suggested by (Byrne, 2013a; Schumacker et al., 2010). The CFA method is employed to examine the factor structure of all influencing characteristics (observed variables), whereas SEM is used to model a network of structural relationships that exist between observed variables and latent variables.

In the beginning, the model is specified as per the results of EFA and hypothetical path model. Pathways are drawn accordingly. To prove the hypothesis, a one-way directional path is connected from II to PI to test the relationship between institutional and pandemic influence. Institutional influence (II) is exogenous, and pandemic influence (PI) is an endogenous variable reliant on II. Overall, the model is constituted by 27 variables that consisted of 12 observed and 15 unobserved variables and is accompanied by 14 exogenous variables and 13 endogenous variables, as displayed by the SEM output. The SEM measurement model that executed CFA through SPSS AMOS is shown in Figure II.

The model that has been identified with a sample size of 922 is overidentified and recursive, with $\chi^2 = 197.218$ and $df = 52$ (>0), suggesting appropriateness for estimating various pathways (Khine, 2013). The sample size of 922 included in this study justified enough sampling adequacy based on Hoelter's critical N displayed in the SEM output (Hoelter, 1983). By selecting the maximum likelihood estimation method (Byrne, 2013a), SPSS AMOS automatically displayed estimations for all relationships with standardized and unstandardized estimates, which are presented in Tables II and III.

**Table II: CFA estimates**

| Choice characteristics (endogenous variables) | $R^2$ | Total effects based on SRW (β) | |
|---|---|---|---|
| | | On account of II | On account of PI |
| Observed Variables | | | |
| C1 | 0.562 | 0.750 | 0.000 |
| C2 | 0.581 | 0.762 | 0.000 |
| C3 | 0.689 | 0.830 | 0.000 |
| C4 | 0.667 | 0.817 | 0.000 |
| C5 | 0.623 | 0.789 | 0.000 |
| C6 | 0.671 | 0.819 | 0.000 |
| C7 | 0.680 | 0.824 | 0.000 |
| C8 | 0.660 | 0.812 | 0.000 |
| C9 | 0.667 | 0.816 | 0.000 |
| C10 | 0.494 | 0.703 | 0.000 |
| C11 | 0.171 | 0.263* | 0.413 |
| C12 | 0.773 | 0.560* | 0.879 |
| Latent Variable | | | |
| PI | 0.406 | 0.637 | 0.000 |

Notes: SRW, standardized regression weights; $R^2$, squared multiple correlations; *, indirect effects.
Source: SPSS AMOS

Referring to Table II, the R2 values for all endogenous variables ranged between 0.406 and 0.689, which indicated moderate (more than 0.50) to substantial (more than 0.75) strength in estimating endogenous variables, as recommended by (Joseph, 2009) (Cohen et al., 2013), except for proximity (C11) ($R^2=0.171$, $β=0.413$), which showed weak estimation strength but adequate estimates (Chin, 1998) as data narrates to unpredictable human behavior. A higher value of standardized estimates (*β*) accumulated on institutional characteristics (C1 to C10) by virtue of institutional influence (II) proved to be a strong estimation. In the case of pandemic influence (PI) ($R^2=0.406$, $β=0.637$), the strength of determination is moderate, with 40.6% of its variance explained on account of institutional influence (II). This means that if II increases by one standardized unit, PI will rise by 0.637 standardized units. Proximity (C11) is explained with 17 percent of its variance on account of PI. It will rise by 0.413 if PI goes up by one standardized unit (direct effect) and will rise by 0.263 standard units if II goes up by one standard unit (indirect effect). On the other hand, 77.3 percent of the variance in suitability under COVID-19 (C12) is estimated by PI. It will increase by 0.879 standardized units if PI goes up by one standardized unit (direct effect) and will increase by 0.560 if II goes up by one standard unit (indirect effect).

For the exogenous component, institutional influence (II) is assembled with 41.7% of its variance (CR=12.039>1.96, $p<0.001$), which is a moderate strength and reasonable value in behavioural research. Referring to Table III, CR values associated with all pathways showing relationships between latent variable (II) and observed variables (C1 to C10) and between latent variable (PI) and observed variable (C11) are above 1.96. This further confirmed that strong convergent validity exists, as all scale items utilized in the CFA model have shown statistically significant loadings in hypothesized directions (Hair et al., 1998). In the case of relationships between two latent variables, II and PI, based on the $B$ value, there is a positive relationship between them, indicating that if II goes up by one unit, then PI will go up by 0.942 units.

**Table III: CFA – Variance and relationships with internal consistency**

| Variance and relationships | $B$ | SE | CR | $p$-value |
|---|---|---|---|---|
| **Component II** | | | | |
| (Variance) | 0.417 | 0.035 | 12.039 | <0.001 |
| (Relationships) | | | | |
| C1←II | 0.988 | 0.045 | 21.873 | <0.001 |
| C2← II | 0.967 | 0.044 | 22.231 | <0.001 |
| C3←II | 1.058 | 0.044 | 24.171 | <0.001 |
| C4← II | 1.062 | 0.045 | 23.789 | <0.001 |
| C5←II | 1.003 | 0.044 | 23.014 | <0.001 |
| C6←II | 1.075 | 0.045 | 23.861 | <0.001 |
| C7←II | 1.063 | 0.044 | 24.010 | <0.001 |
| C8←II | 1.019 | 0.043 | 23.672 | <0.001 |
| C9←II | 1.037 | 0.044 | 23.785 | <0.001 |
| C10←II | 1.000 | -- | -- | -- |
| **Component PI** | | | | |
| (Relationships) | | | | |
| C11←PI | 0.562 | 0.068 | 8.282 | <0.001 |
| C12←PI | 1.000 | -- | -- | -- |
| **Hypothesis** | | | | |
| PI←II | **0.942** | **0.057** | **16.434** | **<0.001** |

Notes: Relationship: observed variable and latent variable, $B$, regression weights; SE, standard error; CR, critical ratio.
Source: SPSS AMOS

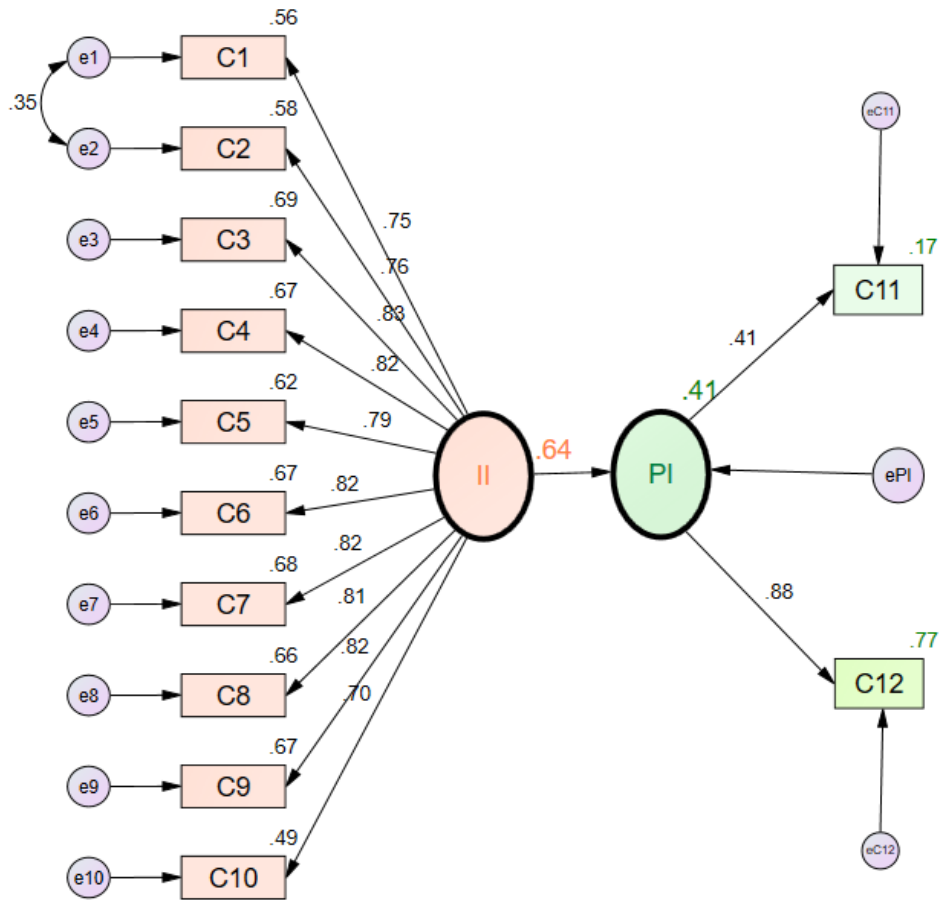

Figure II: Structural equation modelling on 'choice characteristics during pandemic'
Source: SPSS AMOS

*Model fitness and Hypothesis validation*

Fitness indices obtained for the measurement model of this study are noticed in accordance with various fitness indices recommended for SEM and hence support the plausibility of the relations among variables (Teo, 2013) (refer Table IV).

The research hypothesis of this study that states there is no significant relationship between institutional influencing characteristics and pandemic influencing characteristics involved in the choice decision of EIs under the COVID-19 pandemic situation is tested by knowing the relationship II←SI (refer to Table III and Figure II), which shows that this relationship is statistically significant in the positive direction (*B*=0.942, CR=16.434, *p*<0.001). A one-unit rise in II will result in a 0.942-unit increase in PI. Hence, null hypothesis $H_0$ is rejected, and alternative hypothesis H is accepted.

**Table IV: Fitness of model**

| Fitness indices | Recommendable limits for model | Measurement model under study | Literature support | Interpretation about Model fitness |
|---|---|---|---|---|
| $\chi^2$ | Insignificant for N<250 | 197.218 (Significant for N=922) | (Anderson & Gerbing, 1988) | Good fit |
| Ratio $\chi^2/df$ | <5 for (N>500) | 3.793 (N=922) | (Marsh et al., 1988) (Marsh & Hocevar, 1985) | Good fit |
| Hoelter's critical N | N=368 (minimum) for $p<0.001$ | N=922 | (Hoelter, 1983) | Good fit |
| TLI | >0.95 | 0.975 | (Tucker & Lewis, 1973) | Good fit |
| CFI | >0.95 | 0.980 | (Bentler, 1990) | Good fit |
| RMSEA | <0.5 ($p$ of close fit >0.05) | 0.55 ($p$ of close fit >0.05) Fit of model is close | (Hu & Bentler, 1999) | Good fit |

All scales under study are empirically tested for reliability and validity using both EFA and CFA. The SEM model has successfully presented a combination of a hypothetical path model and a CFA model that statistically answered the research questions and validated the research hypothesis of this study; henceforth, the research objective is achieved here. By comparing the indices required for good fit (refer to Table IV), the model – 'choice influencing characteristics under the COVID-19 pandemic situation' – achieved a good fit, as specified below.

$\chi2(52, N=922) = 197.218, p< 0.001$, CFI = 0.980, TLI = 0.975, RMSEA = 0.055 (CI90 0.047, 0.063, $p=0.146>0.05$)

The model has thus demonstrated that the performance of the concept appears to be stable and robust, with all relationships that are hypothesized to be measuring what this study has set out to evaluate.

**Statistical inference and discussions**

This study has verified the influence of choice characteristics associated with EIs and the COVID-19 pandemic in regard to the selection of an EI during the COVID-19 pandemic situation. It has also verified that a relationship between institutional characteristics and pandemic characteristics arises due to the COVID-19 situation. Despite the fact that the performance of institutional influencing characteristics in pandemic situations is as usual as that in ordinary situations, it incredibly has affected pandemic influencing characteristics through proximity to the hometown and suitability under COVID-19.

During the pandemic, institutional influence ($\beta=0.417$) is significantly accumulated by the usual institutional choice characteristics. This was also evidenced by several studies, as discussed below, when the situation was nonpandemic.

The importance of location and locality ($\mu=3.992, R^2=0.562, \beta=0.750, B=0.988$) in making EI choices is evidenced by this study. During the COVID-19 pandemic, the 'infected area' related to coronavirus was the key anxiety for students; hence, they assessed it in terms of its spaciousness, airy ventilation, accessibility and suitability of facilities and amenities wherein it was situated. Similar findings were stated under nonpandemic conditions by (Sovansophal, 2019), who showed that a good location and locality are constructive in fetching enrolments on campuses.

Trust and beliefs are the key dimensions of image and reputation (Finch et al., 2013). During the pandemic, when almost nobody is aware of EI performance, students have no other options but to rely on EIs to provide suitable crisis management practices (Maringe & Gibbs, 2009) for continuing pedagogy that mitigate the risk of COVID-19. Furthermore, as the buying behaviour of customers in a pandemic crisis is believed to be a function of organizational reputation and trust (Coombs, 1998), EIs with a good image and reputation are more likely to be trusted under the COVID-19 situation. Because of this, students in this study perceived image and reputation as an important characteristic (μ=4.120, $R^2$=0.581, $β$=0.762, B=0.967) in selecting their EIs.

Faculty acts as facilitators and mentors in preparing, interacting and motivating students to achieve their academic goals (Salami, 2007). Their support and motivation can be vital for students' emotions that students desperately require during the pandemic to improve their distress for better psychological well-being (Sood & Sharma, 2021). This is why the faculty profile (μ=3.964, $R^2$=0.689, $β$=0.830, B=1.058), as usual, is treated as an important influencing characteristic that facilitates the choice of EIs. (Bao, 2020) documented similar conclusions in terms of the importance of faculty assistance in impacting and sustaining higher education during the COVID-19 pandemic period.

Alumni status is another causative characteristic of EIs, vital for potential students and their family in making their EI choice. Alumni's overall status, such as their reputation gained after graduation (Ho & Hung, 2008) and their employment position (Kalimullin & Dobrotvorskaya, 2016), holds significance during pandemic situations, as students can analyse the risk involved in enrolling in perticular EI against the benefits the students receice after their graduation. This appeared to be true in this case, as students valued alumni (μ=3.906, $R^2$=0.667, $β$=0.817, B=1.062) in making their EI choice.

As the majority of entry-level jobs in the engineering profession during the pandemic are diminishing, campus placements can only provide students with a breakthrough that makes their engineering career worthwhile. During the pandemic, campus placement activities of EIs can offer rewarding benefits offered in terms of skill development that make students competitive in the world and offer better employment opportunities in job crisis situations during the pandemic. This is what students under this study might have perceived and hence campus placement (μ=3.979, $R^2$=0.623, $β$=0.789, B=1.003) of EIs is proven to be a governing character in deciding EI choice, which is evinced by (Matusovich et al., 2020a).

This study has revealed that quality education (μ=3.937, $R^2$=0.671, $β$=0.819, B=1.075) is an important institutional characteristic in deciding EI choice. The notion of 'quality' in higher education is a function of tangible facilities, intangible services and human relations. Students under this study acknowledged its importance in delivering an excellent learning atmosphere during the COVID-19 pandemic situation, and the need for such an atmosphere was also noted by (Zuhairi et al., 2020).

Infrastructure and facilities (μ=3.911, $R^2$=0.680, $β$=0.824, B=1.063) is a backbone and fundamental support system of the higher education system that needs to be rendered through its suitability, accessibility and affordability to continue pedagogy during pandemic situations (Raaper & Brown, 2020). Hence, the students under this study were influenced in making their EI choice.

During pandemic situations, preventive measures and following mandatory standards and guidelines (UGC, 2021a) are the only ways to deliver pedagogy to students' overall wellbeing (Cheng et al., 2020). Today, safety and secured arrangements are contemplated as personal protection shields for students during pandemic situations. For this reason, this study has observed safety and security (μ=3.972, $R^2$=0.660, $β$=0.812, B=1.019) as a key influencing characteristic in making EI choices.

Curriculum delivery during pandemics is the most difficult challenge for engineering studies, and redesigning it via online, onsite or hybrid modes in pandemic situations is an urgent need (Cahapay, 2020) that reduces the burden of cost and workload and eases mental stress. In pandemic situations, successful curriculum delivery is entitled with gaining, accessing, and practicing knowledge and skills building that keeps students' interest live by inculcating proper social distancing. Therefore, curriculum delivery (μ=3.929, $R^2$=0.667, $β$=0.816, B=1.037) is a key characteristic of EIs that influences students in choosing their EIs.

During pandemic situations, cost-effectiveness, convenience, time, and efforts spent are more vital, as they relate directly to the mental and health conditions of students. For this cause, value for money (μ=3.764, $R^2$=0.494, $β$=0.703) has a positive influence in directing students' decision making.

Referring to proximity (μ=3.329, $R^2$=0.171, $β$=0.413, 0.263, B=0.562), this study has indicated that it has impacted creating pandemic influence though low influence but does affect students' choice. The lower ratings on the importance than other choice characters might be because EIs are generally situated at a distance that is far from the hometown of students, and they might have no other option but to select it. However, it is controlled by pandemic as well as institutional influence in a positive direction. This means that if pandemic influence increases, proximity also increases. This means that students will choose EI, which is nearer to their hometown, as this decreases the distance travelled, saves time and cost for the family, and sustains health-related safety and security during the pandemic. This further justified that EIs situated near students' markets are better in position to be selected by local students (Matusovich et al., 2020b), particularly in the pandemic situation. This also supports the findings of (Mok et al., 2021), who realized that institutions that are placed at a far distance have more to work on reframing policies to make them suitable for students during the pandemic to attract them.

The influence of suitability under the COVID-19 pandemic was employed for the first time in this study. It denotes an environment that brings normality into engineering pedagogy with the ease of accessibility and flexibility that is appropriate under the COVID-19 situation by following social distancing standards. It ($\mu$=3.502, $R^2$=0.773, $\beta$=0.879, 0.560) is a major outcome of pandemic influence and is affected by both pandemic influence and institutional influence, which is helpful in determining institute choice.

Last, pandemic influence is well governed under the impact of institutional influence. It is thus confirmed that traditional choice characteristics strongly direct students' perceptions about institutional standings in crisis situations such as COVID-19 and its suitability under pandemic conditions. Hence, it is recommended that EIs reposition themselves to be perceived as suitable under the COVID-19 pandemic situation.

**Implications, suggestions, and contribution**

According to the findings of this study, along with the consideration of pandemic influence, traditional institutional influencing characteristics must be reconsidered to enhance suitability under pandemic conditions. During the pandemic, institutional characteristics seem to have strong and positive impressions on perceptions of pandemic influence, including suitability under COVID-19, which enhances institutional choice. Thus, this study has explored how existing institutional characteristics can control situational influence. The following managerial implications and suggestions are envisioned for the effective performance of EIs during the pandemic by reframing institutional characteristics.

During the pandemic, institutional governance and best practices involving quality education, care taking faculty, students centric facilities and suitable curriculum delivery that keeps the interest of students ongoing, minimizes their academic loss, creates a feeling of being affiliated and justifies them as ethical engineers are very important in developing a high prestige and high reputation of EIs (Gill et al., 2018). Furthermore, incorporating quality infrastructure and facilities along with effective crisis management measures (Maringe & Gibbs, 2009) during the pandemic will trigger positive insights into the quality of EIs (Hemsley-Brown & Oplatka, 2015a).

With one action, it has two-fold benefits for EI during the pandemic. First, providing quality education and services will positively improve image and reputation (Khoi et al., 2019). Second, it will build trust in EIs' commitments to providing quality services. It will achieve students' reliability and confidence in quality provisions rendered by EIs during the pandemic. EIs further need to create an indorse co-creating mechanism for providing and processing vital information about their offers for informed choice decisions (Mogaji & Maringe, 2020). EI stakeholders, such as faculty and existing and alumni students, are real experience holders and direct sources of spreading 'word-of-mouth' about 'suitability' during the pandemic.

Due to the immobility of institutions' physical assets, EI has little to work in proximity. However, as this study has predicted the importance of proximity to the hometown, it becomes binding on local institutions to provide excellent educational services with social distancing norms to grab new enrollments. The EIs must realize that 'all size does not fit all'. EIs should inhouse all required facilities that meet diverse expectations under the COVID-19 pandemic. The success of EIs will be dependent on how far it creates a 'house of reliance' (Nandy et al., 2021) for them. All such efforts will ultimately develop institutional image (Manzoor et al., 2020) and long-lasting relationships (Clark et al., 2017), which is the need for hours foreseen in creating future markets for EIs during pandemic situations.

Nevertheless, EIs should process their repositioning by following pandemic guidelines issued by government and apex authorities (UGC, 2021b) from time to time. If the pandemic carries with us for a long life, then the institute will have to open up other options, such as small campuses and relocation in remote places (Gross, 2020).

To the best of our knowledge, this study is the first to present insights into the performance of choice characteristics during the COVID-19 pandemic that are utilized to select an engineering institute during the pandemic situation (2020-2021). Guessing is, it is also first to come up with new look-out 'pandemic influence', which is noticed to be a significant utility in evaluating choice characteristics under pandemic conditions. It has successfully examined and explored the relationship of suitability and proximity with the influence of pandemics as well as traditional institutional characteristics. Next, it has come with significant evidence that traditional institutional influencing characteristics associated with EIs are positively related to pandemic influence. This is the main contribution of this study.

The study has arrived with substantial hopes for academicians and policy makers. As it has firmly established and deeply rooted in most challenging task: administering new enrollments. EIs will have to reposition themselves to normalize pandemic influence by tuning institutional characteristics. EIs and aspiring students will be known about how choice decisions are influenced under pandemic conditions. The SEM model of this study can be a yardstick for EIs to stay ahead in competitive markets. In the future, if the pandemic continues to be with us for a long time, this study is highly supportive of its revolutionary road, which is visible and feasible for bringing future students into EI campuses. Accordingly, the study has added new and substantial materials and thus has made several key contributions to the existing body of knowledge.

**Conclusion**

COVID-19 has impacted the higher education sector globally, including Indian EIs. It tightened its knot around EIs, which forced them to bring their previous half-shut shades completely down. EIs are now at more risk while doing nothing during the pandemic. Potential students live and grow with their life-dream 'institute going'. Regardless, the mindset of both EIs and potential students should be tailored to 'show must go'. Fetching new admissions to EI campuses before the pandemic was a difficult task, and during the pandemic, it became a 'now or never' situation for EIs. However, the current study analytically mapped the influence of choice characteristics that regulate pandemic influence and are useful in choice decisions under pandemic conditions.

The main objective of this study was to examine the influence of choice characteristics and, consequently, to critically explore relationships of institutional influence and pandemic influence during the COVID-19 pandemic. Research questions were qualitatively answered, and the associated hypothesis was statistically validated. First, the study has noticed that traditional institutional characteristics governing choice decisions also have a predominant influence in pandemic situations. Next, the results have confirmed that the proximity and suitability of EIs under pandemic conditions are the key characteristics that are statistically and positively linked to pandemic influence. Specifically, the findings exposed a positive relationship between various traditional institutional influences and pandemic influences, where institutional influence strongly commands pandemic influence.

To culminate, at this moment, it is dubious that EIs will be weathering a 'new normality' during the pandemic. The answers to this question are very reliant on EI's resilience in reframing student-centric practices that govern suitability under pandemic conditions. For the moment, it is time to 'change for better' in the form of tangible and intangible provisions that intensifies demand for engineering education and expediting choice decisions during the pandemic. This evolution may bring 'normality' to 'new' enrollments and can become a revolutionary transformation in the future ahead!

**Limitations and future research**

Like any research that employs a limited sample, this study is restricted to the fact that it deals with a single context, the North Maharashtra region of India, so that its findings cannot be directly generalized. All things considered, the current study's sincerity and relevance lies in exploring the relationship of 'pandemic influence' with traditional influencing characteristics. Realizing these facts, plenty of research doors are open to investigating the influence of institutions offering study in other disciplines in different regions. Such future studies may report various relationships, as choice characteristics and pandemic impact vary with the region wherein the institutions are situated; consequently, various perspectives on pandemic influence and suitability of the institutions can be acquired under pandemic conditions.

Next, the survey was conducted during the COVID-19 pandemic, and the findings may not be similar in a normal situation. Another fact is that the choice process for students basically begins in their precollege days. In India, as this pandemic has arrived in 2020, some students may not have much exposure to its influence. Henceforth, future research is encouraged periodically but frequently that includes a choice process over the entire pandemic period. Pandemic influence and suitability under COVID-19 are utilized for the first time in this study to give general ideas about their relationship with choice characteristics in selecting EIs under pandemic conditions. Although sufficient progress on choice characteristics has been escorted in the first attempt, a more refined and detailed scale can be developed in future research.